# Improvement in Cell Imaging by Applying a New Natural Dye from Beet Root Extraction

Arpita Das[1], Debarati De[1], Ajay Ghosh[1], Madhuri Mandal Goswami[2]*

[1]CRNN division, Calcutta University, Block-JD, Sector III-2, Salt Lake, Kolkata-700106, India

*[2]S. N. Bose National Centre for Basic Sciences, Block-JD, Sector III, Salt Lake, Kolkata-700106, India

*Email: madhuri@bose.res.in

Fluorescent dyes are getting popularity for last few decades due to their extraordinary applications in cell imaging. We have discovered organic Beet root extracted fluorescent (BREF) dye as an efficient pigment for effective cell imaging. By applying this dye to different types of human cells we obtained very good results in fluorescence cell imaging in case of all types of cells. We also have seen that this dye takes very less internalization time to give very good image of cells and remains stable for sufficiently long period.

Use of the fluorescent molecules for bio imaging and biomedical applications has attracted considerable attention over the past few decades. People are using different types of fluorescent molecules such as fluorescent proteins[1], highly conjugated polymers[2], fluorescent inorganic materials[3], dye[4] etc . However, most of them show many disadvantages, which many times limit their practical applications in imaging purposes and they are not at all cost effective. Most of the organic dyes are hydrophobic and unstable in biological media. Sometimes these types of molecules show photo-bleaching property during prolonged experimental period with yield of very low molar absorbivity. Quantum dots with high yield fluorescence intensity are toxic to living organisms due to their heavy metal toxicity and inorganic fluorescent materials are non biodegradable[5]. Because of all these above mentioned disadvantages people are in search of some new organic fluorescent probes (OFP), which should be nature friendly. Scientists are giving much effort to discover new dyes for fluorescence imaging to investigate and understand the phenomena more precisely[6-18]. Different types of biological phenomena can be investigated by tagging or labelling the bio-molecules present in different parts of cells and it has become the most applicable and effective tool to observe the structural information and intra-intermolecular interaction within the biological systems[19-21]. We synthesize magnetic NPs and investigate their properties for different biological applications such as cell imaging, hyperthermia therapy, drug delivery and release[22-24] etc. For such types of use, the NPs must be selectively attached with cells and attachment should be probed for further investigation. Investigation becomes very authenticate and clear if probing can be done by some visualization technique. The attachment of NPs with cells can be done sure by probing them with the help of fluorescence imaging technique. For this, NPs are tagged with different fluorescent dye molecules and then this conjugated molecule is attached with cells. So we came up with the idea of using an organic dye which will be very easy to manufacture & handling and very intense fluorescence image should be given after attachment with cells in presence of different magnetic nanoparticles. We have extracted a dye from the liquid part of BEET ROOT pulp. After removing the liquid part from the beat root pulp, this part directly can be used as fluorescent probe for cell imaging with no further processing of the dye. We have tagged this beet root extracted dye with sqamous epithelial cell line and are able to have cell imaging result after 10 or 15 minutes of staining. Hence the molecules are tagged with the cells very quickly and efficiently. The results are very much sustaining and reproducible also. The major advantages of using this dye is it is nontoxic, cost effective, it gives very prominent fluorescent images in red and green light both compared to other available dye.

Experimental procedure

Materials and characterization   Beet root from market, Ferric chloride hexahydrate ($FeCl_3.6H_2O$), Cobalt chloride hexahydrate ($CoCl_2.6H_2O$) and alcohol histopaque-1077 were purchased from Sigma Aldrich. Potassium hydroxide (KOH), phosphate buffer solution (PBS), Dimethyl sulfoxide (DMSO), ethanols were purchased from Merck, Germany. Culture media Dulbecco's Modified Eagle's Medium (DMEM), non essential amino acids, Amphotericin B, RPMI-1640, penicillin, streptomycin, gentamycin, L-glutamine and were purchased from HIMEDIA (Mumbai, India). Fetal bovine serum (FBS) was purchased from Invitrogen (Carlsbad, CA, USA).

Labelling magnetic nanoparicles with beet root extracted dye At the beginning 25 mg of cobalt ferrite nanoparticle ware mixed with 2ml 0.1 M $NaHCO_3$ solutions and probe-sonicated to disperse the particles well. 50 µl of beet root extracted concentrated fluorescence dye was dissolved in 2 ml of aqueous dimethyl sulfoxide (DMSO) and mixed instantaneously to the prior cobalt ferrite solution. Then the final solution was set in the rocker at room temperature for the next 24 h in the dark. Then the final solution was centrifuged to separate cobalt ferrite particles and washed with water to remove excess unbound dye.

Application procedure of this fluorescence dye for cell imaging

Breast cancer cell imaging For breast cancer cell imaging we have taken MDAMB-231 (breast cancer) cell line. Cell imaging was done after incubation of cells with beet root extracted fluorescent (BREF) dye functionalized particles and with only BREF dye.  First 1 mg of cobalt ferrite particles functionalized with BREF dye was dissolved in 1 ml of de-ionized water and sonicated to disperse the particles for further use. The cancer cells were seeded in 12 wells plate and incubated overnight. After that the cells were further incubated for 24 hrs to check the interaction of cells with dye functionalized particles in first 6 wells and with dye except particles in 2[nd] 6 wells. The experimental protocol was fixed by addition of dye fuctionalized particles in different doses (10-50 µl/well) in first 6 wells where one well of first row of the first 6 wells was considered as control. In the next 2[nd] set of 6 wells the order was maintained same as in 1st 6 wells but the differences was, here the direct beet root extracted dye solution (10 µl of BREFdye was dissolved in 1 ml of deionized water without attaching with the NPs ) was added taking first one well as control and next 5 wells treating with different doses of dye (10-50 µl/well).

Normal cell imaging

Peripheral blood mononuclear cell (PBMC) isolation Normal cell imaging was done on peripheral blood mononuclear cells (PBMCs). PBMCs are isolated from human blood after collecting the blood in heparinised vacutainer blood collection tubes (BD, Franklin Lakes, NJ). The blood was added very slowly in a layer into a centrifuge tube containing 120 ml of histopaque 1077 with an angle of $45^0$ so that the blood is not mixed with histopaque, which was then centrifuged at 3000 rpm for 30 min. Lastly the PBMC layer is collected and washed with phosphate buffer solution (PBS). For seeding the cells, the RPMI- 1640 medium was used with 10% FBS. The cells were cultured in 12 wells plate.

PBMC cell imaging For PBMC imaging purpose, same protocol was followed as was maintained in case of cancer cell imaging which is mentioned above.

Squamous epithelium cell imaging Normal squamous epithelium cells were collected from a healthy person and directly taken in two 1ml petridishes. 20 µl of the above mentioned BREF dye solution was added in these two petridishes and were kept in incubator at $37^0$C for 10 and 15 mins respectively. After 10 and 15 mins of incubation the fluorescence images were taken. This experiment was done to confirm the internalization time of this dye with cell is very less.

3-(4,5- dimethylthiazol-2-yl)-2,5diphenyltetrazolim bromide (MTT) assay To check the cytotoxicity effect of BREF dye, MTT assay has been performed with the cancer and normal cells. The cells are incubated with varying amount of dye (10-50 µl/well) for 24h at $37^0$c in 5% $CO_2$. After that cells are washed with PBS and then further incubation is done in PBS with addition of MTT. After 4hr of incubation there is a cleavage of MTT reagent by viable cells and formazan is formed. To dissole the formazan crystals, the cells are incubated for 30 min with 50 ml of DMSO after removing the medium. Then these coloured solutions are allowed for absorbance measurement by UV-VIS spectrometer at 575nm wavelength and the results are compared with the control.

Absorption and emission studies The absorption and emission spectra were taken with the help of the UV-Vis and fluorescent spectrophotometer for different concentrations of BREF dye with varring the amount from 10 to 70 µl each in 3ml of water and taking the solution in a quartz cuvette.

Result and discussion

Imaging of various type of cells The BREF dye has been applied to different types of cells, like breast cancer cell line(MDAMB-231), WBC, leukemia, Squamous epithelium. All types of transmittance, fluorescnce and merged images for all types of cells are taken and shown in Figure 1. The smear of BREF dye functionalized cobalt ferrite NP are drawn on a glass slide and transmittance, flurescnce(RFP) images are taken then both type of images are merged which are shown in figure 1 (a,b,c). From figure 1 (b) we see that BREFdye fuctionalized NPs give very good fluorescent image.

Particles are washed with water after functionalization by the dye to remove the excess amount of unbounded dye. Hence from this fluorescnce image it is confirmed that the dye is succesfuly attached with the particles.The next six images (figure 1 (d,e,f,g,h,i)) represent the fluorescnce imaging for the normal WBC in two different excitation regions (red(RFP) and green(GFP)). Figure 1(j,k,l) are representing the similar types of images (transmitance, fluorescnce, merged) for squamous epithelium cells and figure 1(m,n,o,p,q) shows the fluorescence imaging on breast cancer cell line upon applying the BREF dye. The last three images (figures1 (r,s,t)) representing fluorescnce imaging on leukemia cells. From all these images it is evident that most types of cell imaging is possible by appying this BREF dye and the image quality also is very good. We have taken the images by normal fluorescnce microscope with low resolution. But if imaging can be done with high end fluorescnce microscope or confocal microscope the individual cell organells might be possible to visualize. From these experiment we can say that this natural BREF dye is very good pigment for cell imaging experiment.

Conclusion In this paper we have shown the staining of the many types of cells with the BREF dye is possible and hence this dye can be used successfully for imaging the cancer and normal cells. We also have shown that the CFNPs are functionalized well by this dye which will help us to monitor the particles by fluorescence imaging technique in case of application of these particles for hyperthermia therapy.

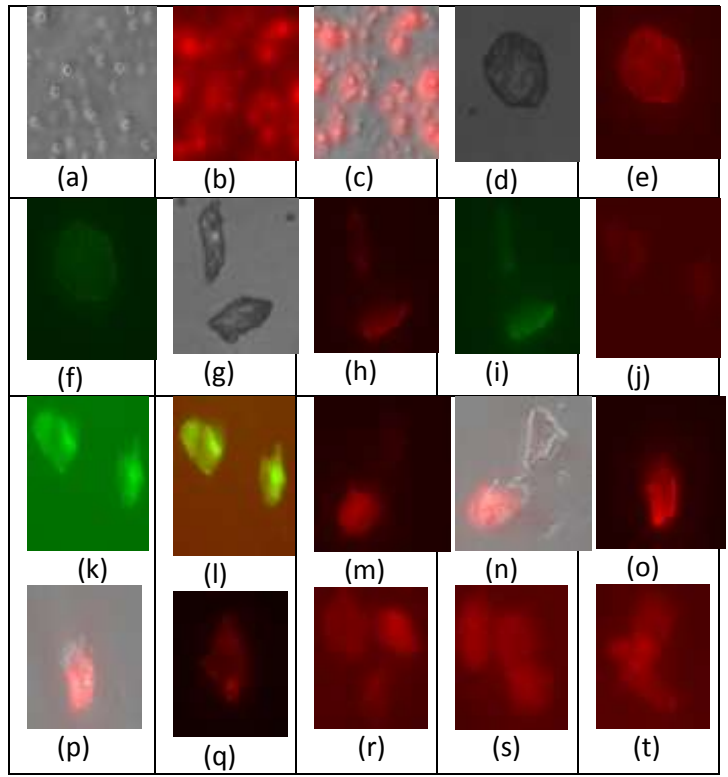

Figure 1

Figure captions:

Figure 1: Different types of cell imaging after tagging them with BREF dye. (a)transmitted,(b)RFP range excited,(c)merge of image (a)and(b), for BREF dye tagged cobalt feritte nanoparticles

(d)transmitted,(e)RFP range excited, (f)GFP range excited,(g)transmitted,(h)RFP range excited (i)GFP range excited images for WBC type cells after staining them with BREF dye

(j)RFP range excited (k)GFP range excited (l)merge of image (j)and(k), of Squamous epithellium cells tagged with BREF dye

(m)RFP range excited, (n)merge of (m) image and its transmitted,(o)RFP range excited, (p) merge of (o) and its transmitted, (q)RFP range excited for MDAMB-231 breast cancer cell stained with BREF dye

RFP range excited three different images (r), (s), (t) of leukemia cell after treating them with BREF dye.